\documentclass[conference]{IEEEtran}
\IEEEoverridecommandlockouts
\pdfoutput=1
\usepackage{cite}
\usepackage{amsmath,amssymb,amsfonts}
\usepackage[ruled, linesnumbered, noend]{algorithm2e}
\usepackage{graphicx}
\usepackage{textcomp}
\usepackage{xcolor}
\usepackage{caption}
\usepackage{enumitem}

\setlist[itemize]{noitemsep, topsep=0pt}

\def\BibTeX{{\rm B\kern-.05em{\sc i\kern-.025em b}\kern-.08em
    T\kern-.1667em\lower.7ex\hbox{E}\kern-.125emX}}
\begin{document}

\title{MMF: Attribute Interpretable Collaborative Filtering}

\author{
    \IEEEauthorblockN{Yixin Su, Sarah Monazam Erfani, Rui Zhang\IEEEauthorrefmark{1}\thanks{\IEEEauthorrefmark{1}Corresponding author.}}
    \IEEEauthorblockA{
        \textit{School of Computing and Information Systems} \\
        \textit{The University of Melbourne}\\
         Melbourne, Australia\\
        yixins1@student.unimelb.edu.au,
        \{sarah.erfani, rui.zhang\}@unimelb.edu.au
    }
}

\maketitle

\begin{abstract}
Collaborative filtering is one of the most popular techniques in designing recommendation systems, and its most representative model, matrix factorization, has been wildly used by researchers and the industry. However, this model suffers from the lack of interpretability and the item cold-start problem, which limit its reliability and practicability. In this paper, we propose an interpretable recommendation model called \textit{Multi-Matrix Factorization (MMF)}, which addresses these two limitations and achieves the state-of-the-art prediction accuracy by exploiting common attributes that are present in different items. In the model, predicted item ratings are regarded as weighted aggregations of attribute ratings generated by the inner product of the user latent vectors and the attribute latent vectors. MMF provides more fine grained analyses than matrix factorization in the following ways: attribute ratings with weights allow the understanding of how much each attribute contributes to the recommendation and hence provide interpretability; the common attributes can act as a link between existing and new items, which solves the item cold-start problem when no rating exists on an item. We evaluate the interpretability of MMF comprehensively, and conduct extensive experiments on real datasets to show that MMF outperforms state-of-the-art baselines in terms of accuracy.
\end{abstract}


\section{Introduction} \label{sec_intro}

In recent years, recommendation systems gain increasing interest by both researchers and the industry \cite{wang2018kdgan, davidson2010youtube}. The most popular recommendation systems are based on collaborative filtering (CF) technique, which provides recommendations based on other similar users' choice \cite{02sharma2013survey}. Matrix factorization (MF) is one of the most common collaborative filtering models, whose main idea is to learn user latent vectors and item latent vectors, so that the inner product of the two vectors can approximate the original matrix with the minimal approximation error. 

MF has advantages of simplicity and performing well in many domains, such as recommendation systems, computer vision and document clustering \cite{liu2017cpmf, wang2018joint, 04shashua2005non, 05xu2003document}. However, it suffers from two limitations. Firstly, its predictions do not provide interpretability, which is showed to be crucial for recommendation systems because interpretations can significantly improve users' acceptance rate \cite{06vlachos2016toward}. Secondly,  if only few ratings are available for some items (a.k.a. item cold-start problem), the prediction accuracy of MF is unsatisfactory. \cite{07lam2008addressing}.

Existing work on enhancing MF mainly focuses on a single aspect, such as improving prediction accuracy or one of the limitations described above. For example, Fuzheng \cite{08zhang2016collaborative} leveraged knowledge information in items to improve the prediction accuracy, but neglect interpretability. Hong \cite{09wang2011collaborative} utilized topic modelling methods to enable matrix factorization interpretability, but the interpretations are limited and the prediction accuracy improvement is not significant.

In order to overcome the two limitations and improve the prediction accuracy simultaneously, we propose a novel recommendation model called \textit{Multi-Matrix Factorization }(MMF). Our model aggregates the \textit{attribute} ratings, where the attributes can be regarded as shared features or characters in items, into item ratings. Specifically, the attribute ratings can be calculated by the inner product of the user latent vector and the attribute latent vector learned through MMF. Further, we integrate two weights to strengthen MMF's performance on both the prediction accuracy and the interpretability, which are the user's preference weight on different attributes and the attribute performance weight on different items. 

\begin{figure*}[h]
\centerline{\includegraphics[width=0.8\textwidth, height=4.9cm]{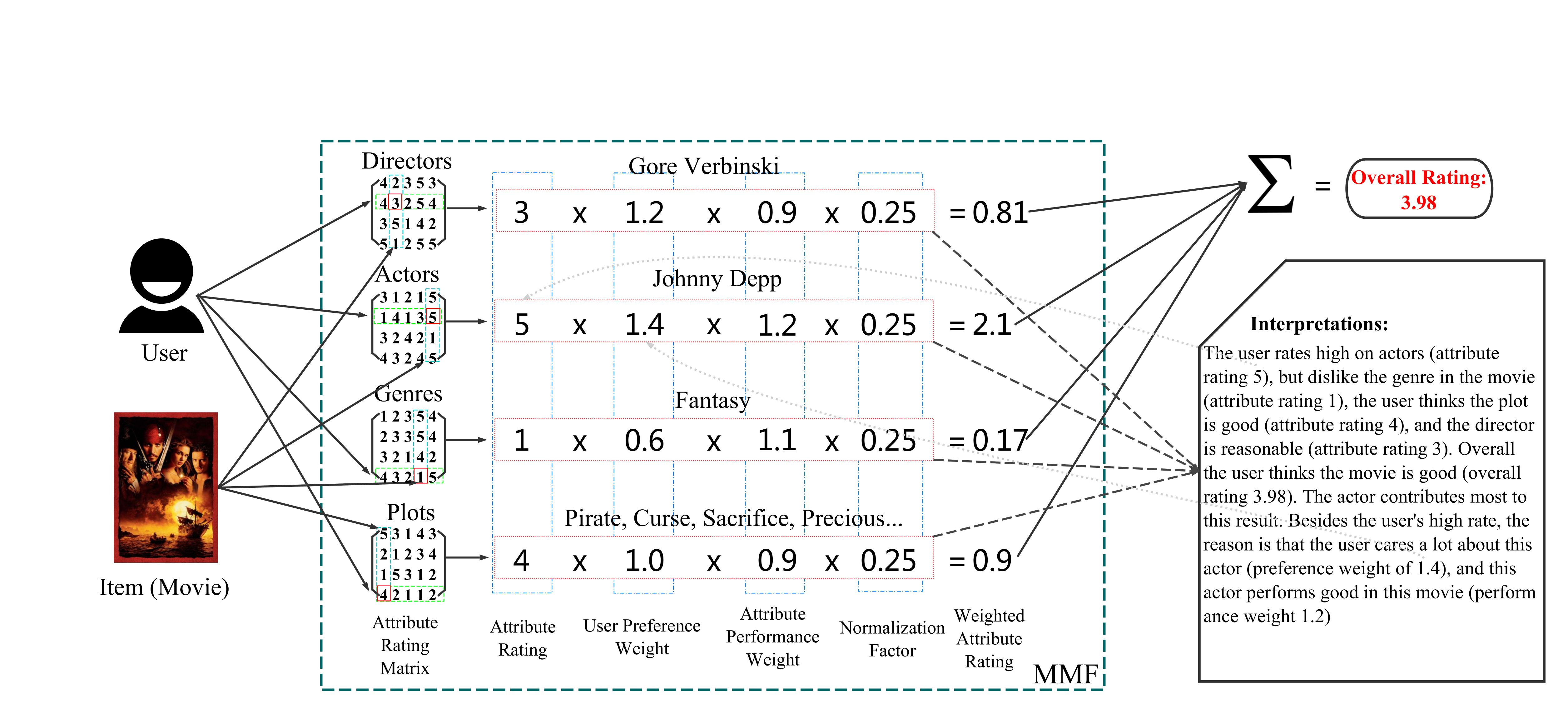}}
\caption{Illustration of MMF's prediction procedure on a movie recommendation example.}
\label{fig_example}
\end{figure*}

Figure \ref{fig_example} illustrates the prediction procedure of MMF on a movie recommendation example. We can see that after training, MMF generates several attribute rating matrices that each matrix represents one type of attribute. Each value in the matrices is an attribute rating a user may give to an attribute. In the prediction procedure, we extract attributes from the movie and find the corresponding attribute ratings. Then, MMF combines the attribute ratings with the user preference and the attribute performance weight together to get the weighted attribute ratings. Finally, the movie rating can be regarded as the sum of all weighted attribute ratings.

The contribution of MMF is that it simultaneously achieves the following goals:
\begin{itemize}[noitemsep,topsep=0pt]
\item \textbf{Interpretability}: MMF explores the contributions of each attribute to the predicted ratings, which enables the model to provide interpretations to the recommendations, such as which attribute in an item may attract a user most. Comprehensive interpretability evaluations are conducted to show the effectiveness of the interpretations.
\item \textbf{Solving item cold-start problem}: MMF infers reasonable predictions in item cold-start situation. Items may share attributes, they are automatically connected by these attributes, which provides the opportunity to transfer attribute ratings from existing items to a new item.
\item \textbf{Better prediction accuracy}: MMF provides a novel way to learn fine grained differences between items, which is useful for improving prediction accuracy. By evaluating on datasets from MovieLens and Netflix Prize, MMF achieves better prediction accuracy in RMSE than state-of-the-art matrix factorization models and attribute based recommendation models.
\end{itemize}

\section{Background and Related Work} \label{sec_background}
In this section, we will discuss several studies related to collaborative filtering with attributes as auxiliary information.

Many researchers leverage attribute information to improve prediction accuracy. Wang et al \cite{wang2015collaborative} proposed Collaborative Deep Learning model that learns item representation from attributes using deep representation model for collaborative filtering. Cheng et al \cite{cheng2016wide} proposed Wide \& Deep model that jointly learn attribute information from wide model and deep model, which combines the advantages of linear models and deep neural networks to achieve better performances. Shan et al \cite{shan2016deep} proposed Deep Crossing that can automatically combines features with the help of stacked Residual Units, which achieves the state-of-the-art prediction accuracy. All of the above models focus on improving prediction accuracy but neglect the interpretability of attributes. 

Some researchers try to enable collaborative filtering interpretability. Chong et al \cite{09wang2011collaborative} perform interpretable article recommendation based on the topics extracted from the articles. They focus on the words extracted for each topic, but neglect other information in articles. Explicit Factor Model (EFM) model \cite{11zhang2014explicit} incorporates collaborative filtering with sentiment analysis on item reviews to realize interpretations. One limitation is that if an item does not have sufficient reviews (cold start situation), the model performs unsatisfactory. Wang et al \cite{wang2018explainable} implement tensor factorization that learns users, items and attributes relationship jointly so that the learned relationships provide personalized interpretations. However, the tensor is very sparse which limits its usage on cold-start situation. Comparing to the above models, our novel MMF model efficiently leverages attribute information so that can simultaneously take into account the interpretability, item cold-start problem and prediction accuracy. 

\section{Our Approach} \label{sec_ourapproach}

We first briefly describe matrix factorization. Then we illustrate the structure of MMF and the learning procedure. Finally, we discuss why MMF is superior to matrix factorization and factorization machine \cite{18rendle:tist2012}, a popular CF model. 

\subsection{Matrix Factorization} \label{sec_mf}
Matrix factorization is one of the most popular collaborative filtering models in recommendation systems. The target of matrix factorization is to learn user latent vectors $U$ and item latent vectors $V$ so that the predicted ratings $R$ can be generated by doing inner product between $U$ and $V$ to approximate real user ratings $X$, which can be written as:
\begin{equation}
r_{ij} = \textbf{u}_{i}^{T}\textbf{v}_{j}, \label{eq1}
\end{equation}
where $i$ is the user index; $j$ is the item index; $r \in R$ is the predicted rating; $\textbf{u} \in U$ is the user latent vector; and $\textbf{v} \in V$ is the item latent vector. 

Then the loss function is:
\begin{equation}
L_{MF} = ||X-R||^{2} + \lambda(||U||^{2}_{F} + ||V||^{2}_{F}), \label{eq2}
\end{equation}
where $||\cdot||^{2}_{F}$ is the Second Normal Form regularization function and $\lambda$ is the scale parameter of the regularization.  

\begin{figure*}[ht]
\centerline{\includegraphics[width=0.80\textwidth]{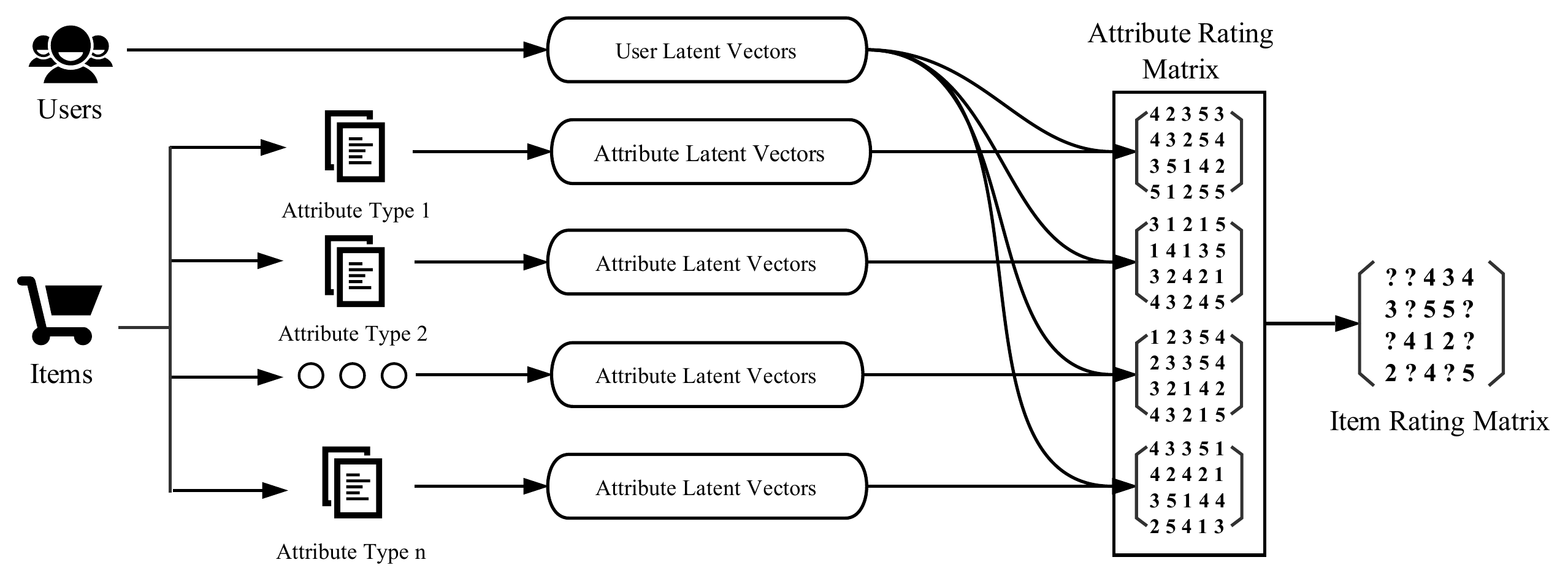}}
\caption{The overall structure of the MMF model.}
\label{fig1}
\end{figure*}

\subsection{Multi-Matrix Factorization (MMF)} \label{sec_mmf}

Towards solving the item cold-start problem and the lack of interpretability of matrix factorization model, we propose the Multi-Matrix Factorization model. The model not only overcomes the aforementioned limitations of matrix factorization, but also achieves better prediction accuracy.

The main idea of MMF is to exploit the common attributes in items. Instead of linking items by user ratings, we build fine grained connections between items using attributes. The MMF model merges several matrix factorization models into a single one. Each matrix factorization model can predict one type of personalized attribute ratings. Here, penalization is achieved by computing a ``preference" weight of a user on each attribute. For example, a user may care greatly about the genre of a movie, while another may care more about the director. Therefore, the final item ratings can be regarded as the weighted sum of the various attribute ratings.

Figure \ref{fig1} shows the overall structure of the MMF model. It first embeds the users and the attributes into user latent vectors and attribute latent vectors, respectively. Then, attribute latent vectors with different types will be trained with user latent vectors in different matrix factorization models to output attribute rating matrices. Each attribute rating matrix stores ratings for one type of attribute. Finally, item ratings can be calculated as weighted combination of attribute ratings. Specifically, users will be represented by user latent vectors $U$, and attributes will be extracted from items and be represented by attribute latent vectors $F$. Therefore, attribute ratings can be calculated as the inner product of the user latent vectors and the attribute latent vectors, or $U^{T}F$.

Next, because the same type of attributes is trained in a single matrix factorization model in MMF, it is inevitable that the attributes with the same type will affect each other, which is unfavourable for the prediction accuracy and the interpretation. For example, if a user likes most genres of movies but strongly dislikes \textit{Thriller}, MMF may not give Thriller a very low attribute rating (which is expected), because the average attribute ratings of all the genres are high. We address this issue by giving each user a preference weight $\omega$ on each attribute to show user's preference on different attributes. Higher $\omega$ value means higher user preference.

Finally, we cannot assume that one attribute performs equally in different items. To better differentiate the quality of one attribute in different items, we assign different weights $\theta$, for attributes in different items, which means that one user may give different ratings for an attribute in different items.

The predicted ratings can be calculated by combining all components discussed above together. In addition, the ratings will be normalized by dividing with the number of attributes in the item to balance inequality of attribute number. The calculation formula is:
\begin{equation}
r_{ij} = \frac{1}{|M_{j}|}\sum_{k \in M_{j}}\omega_{ik}\theta_{jk}\textbf{u}_{i}^{T}\textbf{f}_{k}, \label{eq3}
\end{equation}
where $\textbf{f} \in F$ is the attribute lantent vector; $r$ is the predicted item rating; $M$ is a set of attributes containing in item, and  $|\cdot|$ is the cardinality function. 

\subsection{MMF Loss Function and Updating Algorithm} \label{sec_updatefun}
Although the MMF seems to be more complex than matrix factorization, the training process is not difficult. Next, we will describe how the MMF model is trained.
\subsubsection{Loss Function}
Similar to MF, we use L2 loss function to measure the differences between predicted ratings and user's ratings: 
\begin{equation}
L_{MMF} = ||X - R||^{2} + \lambda(||U||^{2}_{F} + ||F||^{2}_{F}). \label{eq5}
\end{equation}

\subsubsection{Optimization}
In our model, the parameters can be learned by gradient descent methods. Four kinds of parameters in MMF should be learned: user latent vectors $U$, attribute latent vectors $F$, user preference weight $\omega$ and item performance weight $\theta$. Based on the loss function, we can calculate the gradient descent direction for each parameter:
\begin{equation}
\frac{\partial L}{\partial U_{id}}= \sum_{j\in J_{i}}-(x_{ij}-r_{ij})\sum_{k\in M_{j}}\omega_{ik}\theta_{jk}F_{kd} + \lambda U_{id}, \label{eq6}
\end{equation}
\begin{equation}
\frac{\partial L}{\partial F_{kd}}= \sum_{i\in K_{j}}\sum_{j\in J_{i}}-(x_{ij}-r_{ij})\omega_{ik}\theta_{jk}U_{id} + \lambda F_{kd}, \label{eq7}
\end{equation}
\begin{equation}
\frac{\partial L}{\partial \omega_{ik}}= \sum_{j\in J_{i}} (x_{ij}-r_{ij})\sum_{d}\theta_{jk}U_{id}F_{kd}, \label{eq8}
\end{equation}
\begin{equation}
\frac{\partial L}{\partial \theta_{jk}}= \sum_{i\in K_{j}} (x_{ij}-r_{ij})\sum_{d}\omega_{ik}U_{id}F_{kd}, \label{eq9}
\end{equation}
where $J_{i}$ is the items that user $i$ has rated; $K_{j}$ is the users that has rated item $j$; $d$ is the dimension of latent vectors.

\subsection{Relation to Matrix Factorization} \label{sec_whymmf}

MMF improves over matrix factorization in 3 different aspects, as depicted in

\subsubsection{Personalized Interpretable Recommendation}
MMF provides personalized interpretations from two aspects: (i) users' judgment on attributes ($U^{T}F$), (ii) users' personalized preference on each attribute ($\omega$). As Figure \ref{fig1} shown, the item latent vectors can be regarded as the weighted sum of attribute latent vectors. Therefore, items can be divided into several components with explicit meanings. Specifically, $U^{T}F$ can be interpreted as general ratings showing how a user may like a type of attribute, which is similar to matrix factorization on items. $\omega$ will further differentiate the preference of users between different types of attributes. Detailed evaluation will be demonstrated in Section \ref{sec_casestudy}.

\subsubsection{Item Cold-Start Situation}
The main reason that matrix factorization method performs unsatisfactory on item cold-start situation is that if there is no rating history (e.g., a newly added item), the model cannot link it to previous items. MMF overcomes this defect efficiently. An item in MMF is represented as the weighted sum of the attribute latent vectors, who may appear in multiple items so that they can act as the connections. For example, when a new item is added into the system, the item may already contain attributes that appear in previous items. Attribute ratings that users give to previous items can be used as a vital clue to perform personalized item rating predictions. %More specifically, $U^{T}F$ and $\omega$ can be directly used from previous items. $\theta$ is not known for the new item so that it can be assigned as 1 for all attributes. Meanwhile, if some attributes in the new item appear for the first time, we can neglect them and use the attributes that have appeared in existing items. 
Therefore, as long as the new item contains attributes that have appeared in existing items, MMF is able to infer well item rating. %The more items with ratings in datasets, the more attributes appear, then the better the MMF model performs in item cold-start situation.

\subsubsection{Prediction Accuracy}
In addition to the above advantages, attributes extracted from items show the ability on improving prediction accuracy as well. Comparing to matrix factorization, the MMF model analyzes items in a fine grained point of view. It learns the similarities between the attributes rather than the items, which enables the model to learn more delicate differences between items. This helps improve the prediction accuracy. In extreme situations, if there is no common attribute between any two items, MMF is identical to matrix factorization. From Equation \ref{eq3}, we can see that if there is no common attribute, every item latent vector is composed by weights and attribute latent vectors without overlap, which reduces MMF to matrix factorization.

\subsection{Relation to Factorization Machine}
Factorization Machine (FM)\cite{18rendle:tist2012} is a popular attribute-based CF model. FM analyzes attributes by performing inner product among attributes, and the final predictions are the sum of all analysis results, which is similar to MMF. However, instead of aggregating attribute interactions in equal weight, MMF gives attributes interpretable weights (attribute performance weights and user preference weight) to improve both prediction accuracy and interpretability because that the weights can both be parameters to further differentiate attributes' contribution, and can act as informative interpretations (e.g., how attributes perform in different items). We experimentally prove it in Section \ref{sec_ABB}. Meanwhile, MMF simultaneously learns item embeddings (weighted sum of all corresponding latent attribute vectors) during training, while FM does not. This makes MMF easy to further perform item level analysis directly, such as implementing neural collaborative filtering on users and items \cite{he2017neural}. 

\section{Experiments} \label{sec_experiments}
In this section, we conduct extensive experiments to evaluate the performance of MMF empirically. Then, we demonstrate the interpretability of MMF through a case study. Finally, performance of MMF on item cold-start situation and the influence of hyper parameters are evaluated.    

\subsection{Data Description} \label{sec_data_describe}
We use movie recommendation as a representative application in our experimental study due to two reasons. First, there are a few publicly available sources for movie data. Second, movie recommendation has been the focus of previous studies on collaborative filtering such as \cite{ortega2016recommending, ZhengTDZ16}. As future work, we will try to obtain datasets in other domains and excrement on them, too. We use MovieLens \cite{12harper2016movielens} and Netflix Prize \cite{13bennett2007netflix} to evaluate our model, which are commonly used in movie recommendation researches.

\subsubsection{Datasets} \label{sec_dataset}

We use two public datasets: (i) MovieLens 100k (\textit{100k}), (ii) MovieLens latest-small (\textit{20m Light}). Meanwhile, we analyze the performance of MMF on different dataset densities, including rating data density and attribute density (will be illustrated in Section \ref{sec_datastat}). Since the density in public datasets is uncontrollable, we further use three subsets from MovieLens and Netflix Prize Data sources. They are (iii) top movies from MovieLens 20m ranked by box-office (\textit{Boxoffice}) with high rating data density, (iv) \textit{Netflix 500}, and (v) \textit{Netflix 1000} datasets that randomly chose 500 and 1000 movies from Netflix Prize data source respectively, which both have high attribute densities.

In addition, to evaluate our model in item cold-start situation, we randomly pick 10\% movies from all datasets, then regard ratings of the picked movies as test data and the rest as training data, which simulates the item cold-start situation. 

\subsubsection{Attribute Extraction} \label{sec_attrextract}

MMF is based on attributes. Therefore, we extract attributes for datasets. Generally, we can choose any types of attributes. However, to enable MMF the ability of solving item cold-start problem, we pick the inherent attributes of items, which are the attributes that exist when items are generated.

Four types of attributes are concordant to our preference: directors, casts, plots and genres. All attributes are extracted from IMDB website, which is already mapped into the MovieLens data source. We also aggregated Netflix Prize data source to the IMDB database by matching the name and year of the movies. In addition, we used topic modelling method, LDA \cite{15blei2003latent}, to transform poorly structured plot attribute into well-structured form called topics. Each movie will contain one or several topics with the same structure as other attributes. 

\subsubsection{Dataset Statistics} \label{sec_datastat}

Table \ref{tab1} shows the statistic information of all datasets. The \textit{Rate} column demonstrates the rating data density in datasets. The lower the rating data density, the less average rating of users. It is calculated by:
\begin{equation}
 Rate =\frac{N_{r}}{N_{p} \times N_{u}},  \label{sparsity rate}  
\end{equation}
where $N_{r}$ is the number of ratings; $N_{p}$ is the number of movies and $N_{u}$ is the number of users.
\begin{table}[htbp]
\small
\caption{Details of Datasets.}
\begin{center}
\begin{tabular}{|c|c|c|c|c|}
\hline
\textbf{Dataset}&\textbf{Movies}&\textbf{Users}&\multicolumn{2}{|c|}{\textbf{Ratings}} \\
\cline{4-5} 
\textbf{Name} & \textbf{Number}& \textbf{Number}& \textbf{Count} & \textbf{Rate} \\
\hline
100k & 1683 & 944 & 100000 & 0.063 \\
\hline
20m Light & 8788 & 671 & 98688 & 0.017\\
\hline
Boxoffice & 206 & 3550 & 99326 & 0.136 \\
\hline
Netflix 500 & 539 & 5893 & 100181 & 0.031  \\
\hline
Netflix 1000 & 1061 & 17185 & 308544 & 0.017 \\
\hline
\end{tabular}
\label{tab1}
\end{center}
\end{table}

\begin{table}[htbp]
\small
\caption{Attribute details of datasets.}
\begin{center}
\begin{tabular}{|c|c|c|c|c|}
\hline
\textbf{Dataset}&\textbf{Directors}&\textbf{Casts}&\textbf{Genres}& \textbf{Dense Rate} \\
\hline
100k & 1138  & 1988 & 24 & 0.049 \\
\hline
20m Light & 3976  & 11632 & 25 & 0.083 \\
\hline
Boxoffice & 194 & 474 & 20 & 0.050 \\
\hline
Netflix 500 & 28 & 1065 & 24 & 0.103 \\
\hline
Netflix 1000 & 64 & 1837 & 27 & 0.094 \\
\hline
\end{tabular}
\label{tab2}
\end{center}
\end{table}
Table \ref{tab2} demonstrates the detailed attribute information of all datasets. Meanwhile, our experiments generate 50 topics for all datasets. Evaluation on different topic numbers will be demonstrated in Section \ref{sec_topicnum}.  

Moreover, we present an attribute density (\textit{Dense Rate}) to indicate the repetitive degree of attributes in all items. The Dense Rate calculation formula is:
\begin{equation}
 DenseRate = \frac{1}{n}\sum_{i\in \left\{1..n\right\}}  \frac{Ave_{i}}{N_{i}}, \label{eq11}  
\end{equation}
where $n$ is the number of attribute types (in our datasets, $n$ is 4); $Ave_{i}$ is the average number of $i^{th}$ types of attribute that appear in one item; $N_{i}$ is the total number of $i^{th}$ types of attribute in datasets.  

The dense rate is capable of representing the probability that each attribute appears in different items. The higher the attribute dense rate, the higher probability of two items in the dataset share common attributes.

\subsection{Baseline Methods}

Matrix factorization models and State-of-the-art attribute based factorization models are compared, including: 

\renewcommand{\theenumi}{\roman{enumi}}
\begin{enumerate}
\item The \textbf{MF} model that has been discussed before.
\item \textbf{Probabilistic Matrix Factorization (PMF)} \cite{16mnih2008probabilistic}: It uses probabilistic method to describe latent vectors. PMF is one of the most popular variations and usually get better result than MF on large, sparse and imbalance datasets.
\item \textbf{Bayesian Probabilistic Matrix Factorization (BPMF)} \cite{17salakhutdinov2008bayesian}: It uses Bayesian treatment to train PMF with the help of Markov Chain Monte Carlo methods.
\item \textbf{LibFM} \cite{18rendle:tist2012}: It is the official tool to implement FM model. In the experiments, we designed two inputs for LibFM. One only utilizes users' and items' ID as inputs (\textbf{LibFM-s}). Another one includes user, item and attribute information as inputs (\textbf{LibFM-c}).
\item \textbf{DeepCrossing} \cite{shan2016deep}:  It leverages residual deep neural network that automatically combines attributes. We implement this model that stacks three residual units with hidden dimensions 128, 128, 64, respectively.
\end{enumerate}
    
\subsection{Accuracy Results}

We use Root-Mean-Square-Error (RMSE) as the criteria to evaluate the accuracy. Note that the lower the RMSE value, the better the performance. In our experiments, the proportion of training set and test set are 8:2 for all datasets.

\subsubsection{Comparing with Matrix Factorization Methods}

Firstly, MMF are compared with MF and its variations, which are all developed from MF model. The results are shown in Table \ref{tab3}. The best result for each dataset is highlighted in boldface. 

\begin{table}[htbp]
\small
\caption{Performance Comparing with MF and Variations.}
\begin{center}
\begin{tabular}{|c|c|c|c|c|}
\hline
\textbf{Dataset}&\textbf{MF}&\textbf{PMF}&\textbf{BPMF}&\textbf{MMF} \\
\hline
100k & 0.953 & 0.948 & 0.949 & \textbf{0.927} \\
\hline
20m Light & 0.939 & 0.937 & 1.001 & \textbf{0.876} \\
\hline
Boxoffice & 0.761 & 0.760 & 0.794 & \textbf{0.740} \\
\hline
Netflix 500 & 0.971 & 0.967 & 1.043 & \textbf{0.904} \\
\hline
Netflix 1000 & 1.061 & 0.986 & 1.008 & \textbf{0.919} \\
\hline
\end{tabular}
\label{tab3}
\end{center}
\end{table}

From the table, we can see that MMF consistently achieves the best performance over all the datasets. This result suggests that the structure of MMF provides a better way to utilize attribute information and differentiate similar users and items. Meanwhile, we can see that in 20m Light and two Netflix datasets, MMF has more significant accuracy increment than other datasets. According to Table \ref{tab2}, we find that 20m Light and two Netflix datasets have higher attribute density than others. It further indicates that attributes have helped the MMF model to learn fine grained differences between items. In addition, 20m Light and two Netflix datasets have lower rating density than other two datasets according to Table \ref{tab1}, but the accuracy is not influenced by the sparsity of data, which shows that MMF is capable of performing well in sparse datasets. 

\subsubsection{Comparing with Attribute Based Baselines} \label{sec_ABB}

We now compare MMF with the attribute based methods LibFM and DeepCrossing.

\begin{table}[htbp]
\small
\caption{Performance Comparing with Attribute Based methods.}
\begin{center}
\begin{tabular}{|c|c|c|c|c|}
\hline
\textbf{Dataset}&\textbf{LibFM-s}&\textbf{LibFM-c}& \textbf{DeepCrossing}&\textbf{MMF} \\
\hline
100k & 0.993 & 0.938 & 0.935 &\textbf{0.927} \\
\hline
20m Light & 0.896 & 0.889 & 0.880 &\textbf{0.876}\\
\hline
Boxoffice & 0.819 & 0.749 & 0.743 &\textbf{0.740}\\
\hline
Netflix 500 & 0.920 & 0.909 & 0.908 &\textbf{0.904}\\
\hline
Netflix 1000 & 0.936 & 0.925 & 0.921 &\textbf{0.919}\\
\hline
\end{tabular}
\label{tab-libfm}
\end{center}
\end{table}

Table \ref{tab-libfm} shows the RMSE for MMF and baseline methods. From the table, we can see that LibFM-s performs inferior than other models in the table, which again indicates that attributes has positive effect on improving prediction accuracy. Meanwhile, the MMF model is superior in prediction accuracy comparing to the LibFM-c and DeepCrossing on all datasets. Although the increase rate is not as significant as MF methods, MMF shows its advantages on interpretability. Therefore, we can conclude that MMF is not only an interpretable recommendation model, but is also a state-of-the-art attribute based method in terms of accuracy.

\subsection{Different Weight Combinations}

In this subsection, we will analyze how the two weights in MMF, the user preference weight $\omega$, and the item performance weight $\theta$, influence the prediction accuracy. We evaluate 4 combinations: base $U^{T}F$, adding $\omega$ only, adding $\theta$ only, and adding both $\omega$ and $\theta$ (the proposed MMF model).

\begin{table}[htbp]
\small
\caption{RMSE of different weight combinations in MMF.}
\begin{center}
\begin{tabular}{|c|c|c|c|c|}
\hline
\textbf{Dataset}&\textbf{Base}&\textbf{Base+$\omega$}&\textbf{Base+$\theta$}&\textbf{Base+$\omega$+$\theta$(MMF)} \\
\hline
100k & 0.944 & 0.932 & 0.947 & \textbf{0.927} \\
\hline
20m Light & 0.960 & 0.924 & 0.881 & \textbf{0.876}\\
\hline
Boxoffice & 0.759 & 0.757 & 0.747 & \textbf{0.740}\\
\hline
Netflix 500 & 0.912 & 0.914 & 0.904 & \textbf{0.904}\\
\hline
Netflix 1000 & 0.928 & 0.928 & 0.920 & \textbf{0.919}\\
\hline
\end{tabular}
\label{tab-weightcombination}
\end{center}
\end{table}

The prediction results of four combinations are shown in Table \ref{tab-weightcombination}. We can see that the proposed MMF model performs best among all combinations, which shows that the $\omega$ and the $\theta$ together have positive influence on improving accuracy. We can also see that in some situations, such as in 100k and two Netflix datasets, adding one weight cannot guarantee to improve prediction accuracy. This demonstrates that the three components do not work individually in the model, but work together to get the best results.

\subsection{Interpretability Case Study} \label{sec_casestudy}

It is necessary to exam the interpretability of MMF model. In this subsection, we give a case study to validate it in two steps. First, we demonstrate that the MMF model can capture semantic dependencies between attributes. Then, we give an example to show how MMF can provide interpretable recommendations and analyze whether the interpretations are consistent with users' personal interests. 

Specifically,  we choose the movie ``\textit{Guardians of the Galaxy}" from Boxoffice dataset, which containing 8 attributes: 1 director, 1 topic, 3 casts and 3 genres. The topic is the $7^{th}$ topic generated by LDA, which contains the following keywords that indicate the elements in the movie: \textit{superhero, orb, galaxy, musketeer, etc.}. The latent vectors for the 8 attributes will be visualized to show their semantic dependencies, and we will analyze the preference of the 2 users on this movie.   

\subsubsection{Semantic Dependency}

\begin{figure}[h]
\centerline{\includegraphics[width=0.42\textwidth]{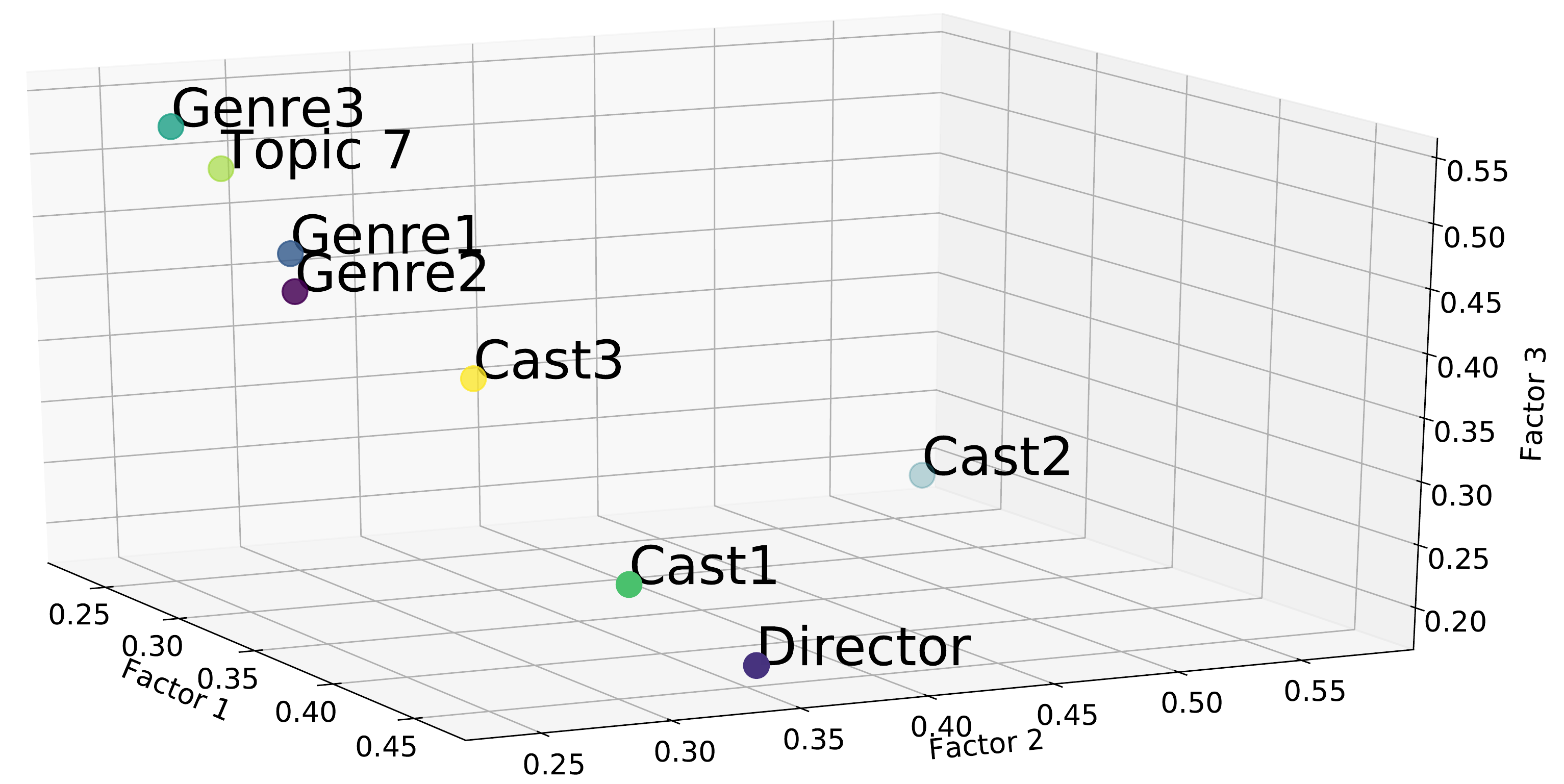}}
\caption{Visualization of attribute latent vectors in one movie (The latent vectors have been compressed into 3 dimensions using t-SNE).}
\label{fig_spatialrelation}
\end{figure}

In this part, we visualize the learned attribute latent vectors in the movie ``\textit{Guardians of the Galaxy}". To draw the latent vectors, we utilize the t-SNE technique \cite{maaten2008visualizing} to reduce the high dimension latent vectors into 3 dimensions while preserving the spatial relationship. Figure \ref{fig_spatialrelation} shows the latent attribute vectors belonging to the movie in a 3-dimension form. Specifically, the attributes are ``James Gunn" (Director), ``Topic 7", ``Chris Pratt" (Cast1), ``Zoe Saldana" (Cast2), ``Dave Bautista" (Cast3), ``Action" (Genre1). ``Adventure" (Genre2) and ``Sci-Fi" (Genre3). 

From the figure, we can see that the attribute latent vectors learned by MMF show strong links between spatial and semantic relationship. For example, attributes ``Action" (Genre1) and ``Adventure" (Genre2) are very close to each other in spatial. Semantically, movies belonging to the genres of both action and adventure account for 23\% of all movies, which is much higher than other genre combinations. Meanwhile, ``Topic 7" is closer to ``Sci-Fi" (Genre3) than another two genres. Intuitively, elements like ``superhero", ``orb", ``galaxy" appears often in movies of genre ``Sci-Fi" than those of genre ``action" and ``adventure", which is constant with our claims. 

\subsubsection{Interests Consistency}

In previous section, we have stated that the MMF model have the ability to recommend items to users with predicted personal attribute preference. It is crucial to show whether the preference predictions are consistent with users' real interests. In this part, we first demonstrate how our model can recommend the movie ``\textit{Guardians of the Galaxy}" to two users with different reasons, and then analyze whether the predictions show the two users' real preferences about this movie based on their rating histories.

\begin{figure}[h]
\centerline{\includegraphics[width=0.42\textwidth]{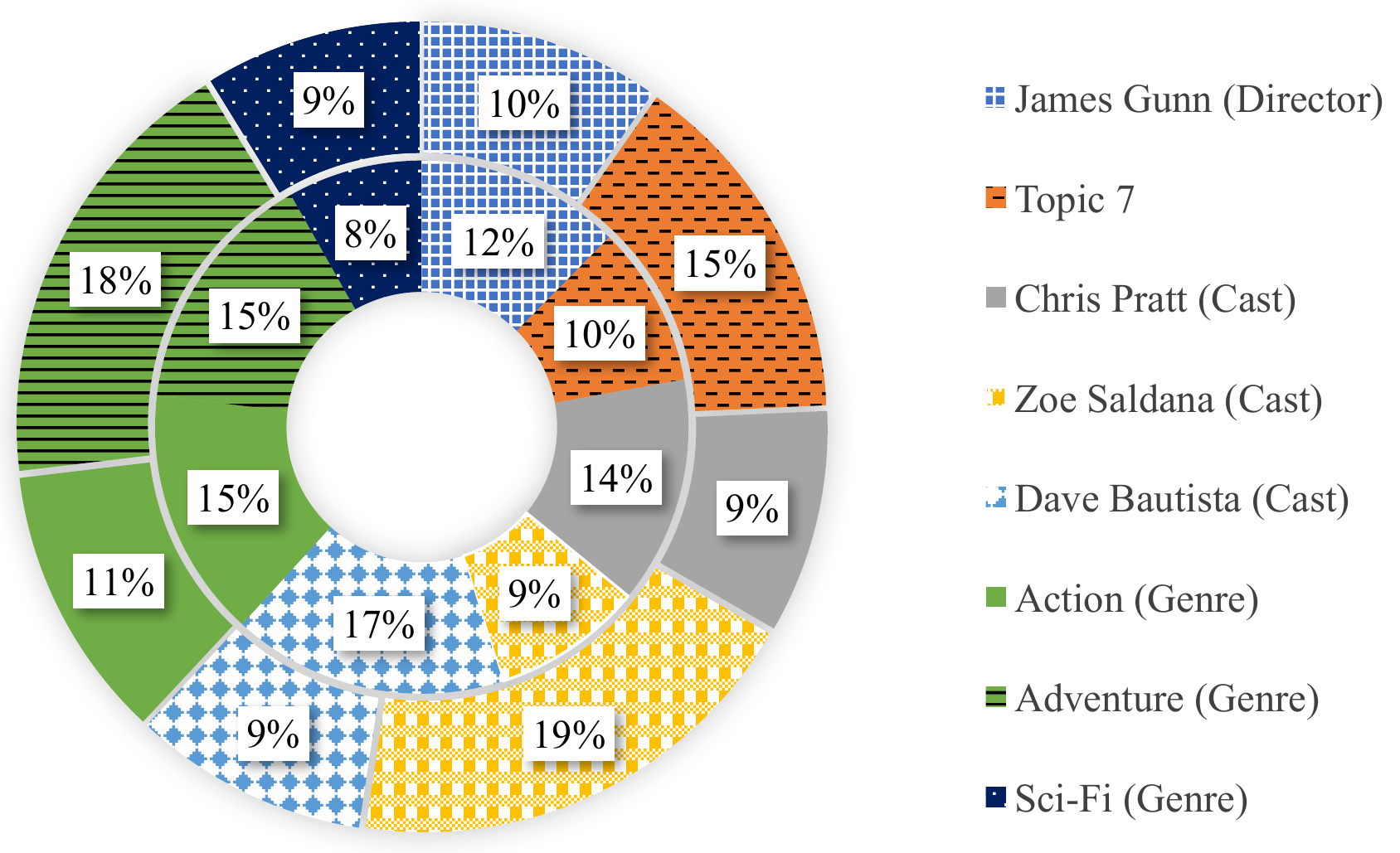}}
\caption{Preference of user A (inner loop) and user B (outer loop).}
\label{fig3}
\end{figure}

In our example, the MMF model predicts that user A and user B will give similar ratings to the movie. In addition, it also predicts that the attribute preference of the two users are different. Figure \ref{fig3} shows the percentage of ratings that compose the overall rating of the movie. The inner loop of the pie chart belongs to user A and the outer loop belongs to user B. We can see that although two users may give similar ratings, the reasons are different.  

From the pie chart, we can see that the rating composition of the two users are different. For example, for user A, the cast Dave Bautista (17\%) may be the most important motivation to watch the movie, while for user B, the cast Zoe Saldana (19\%) and the genre Adventure (18\%) are the attractions. Therefore, MMF can recommend the movie to different users with personalized reasons based on the attribute proportion.

However, our interpretations will be convincing only if they truly capture users' interests. We evaluate it based on an assumption about users' interests: \textit{a user will give higher ratings to movies that contain the attributes the user loves}. This assumption is quantified by calculating an \textit{attribute aware rating difference} (AAD) on each attribute for users. Then we explore whether they are constant with our predictions.

To calculate the AAD of an attribute for a user, we first extract $k$ attributes (in our case study, $k=5$) that are most similar (shortest Euclidean distance) to that attribute, then $R_{k}$ is a set containing all movies that contain these $k+1$ attribute (including the original one) and are also rated by the user. Then the AAD is:
\begin{equation}
AAD = \frac{1}{|R_{k}|} \sum_{j \in R_{k}} r_{j} - \frac{1}{|R_{n}|} \sum_{i \in R_{n}} r_{i},           \label{eq12}
\end{equation}
where $R_{n}$ is a set contains all movies that the user has rated.   

We can see that the first term of Equation \ref{eq12} is the average rating of movies that a user has rated on an attribute (and similar attributes), and the second term is the average ratings of all movies the user has rated. According to the assumption above, if a user really likes an attribute, the AAD of this attribute will be positive, and vise versa. In addition, if a user loves attribute $a_{1}$ more than attribute $a_{2}$, the AAD of  $a_{1}$ is supposed to be greater than $a_{2}$ for this user. Although AAD may not always be accurate due to that a user only rates a small amount of movies, the trend should roughly hold, which means that AAD can reflect users' interests to a certain extent.

\begin{figure}[h]
\centerline{\includegraphics[width=0.40\textwidth]{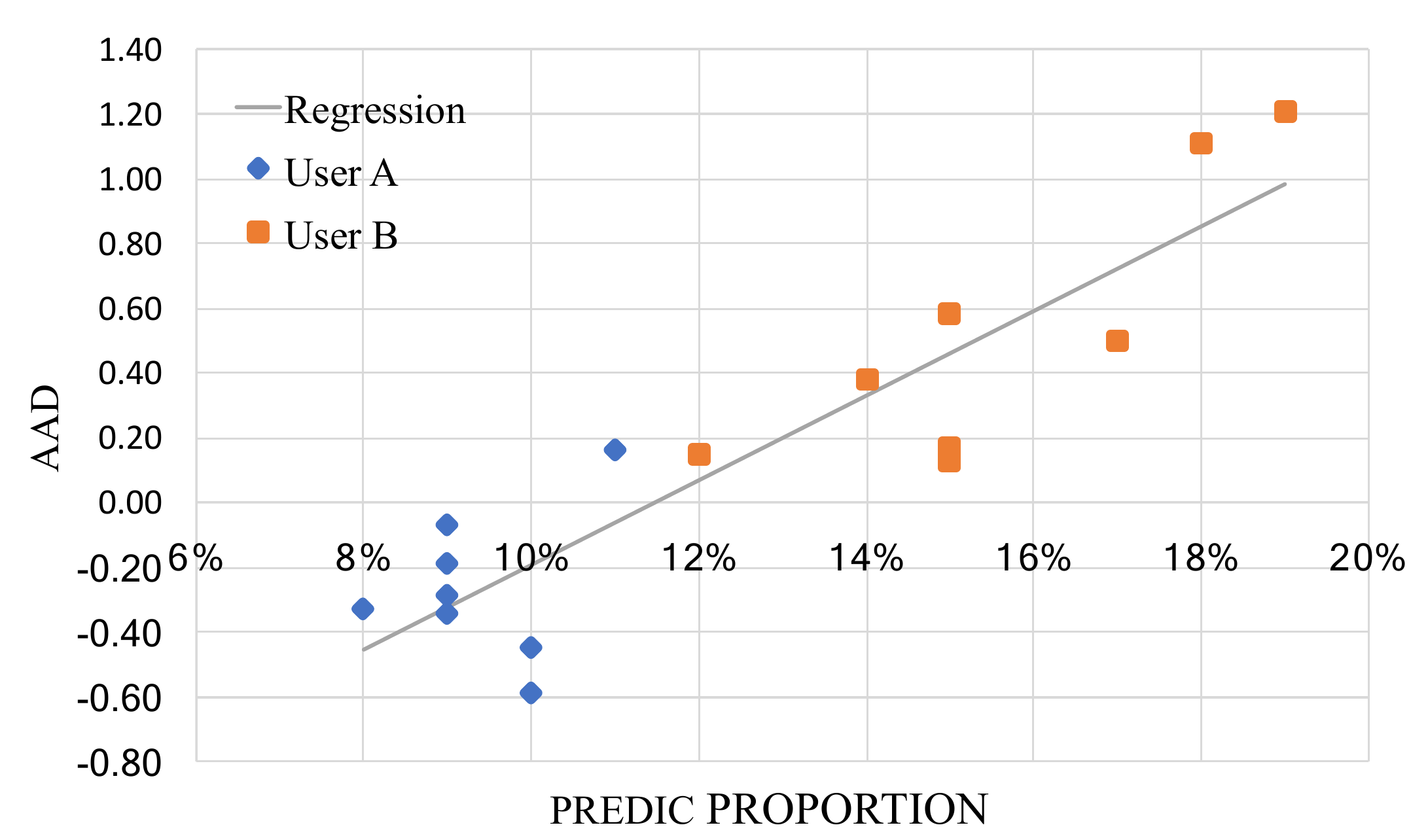}}
\caption{Correlation between predicted attribute proportion and AAD.}
\label{fig_trends}
\end{figure}

Figure \ref{fig_trends} shows the relationship between AAD and the predicted proportion. We find a strong positive correlation between them, which shows that the predicted proportions are consistent with users' interests. Therefore, we can conclude that our recommendations can capture users' interests and the interpretations of the MMF model are convincing.

\subsection{Item Cold-Start Problem}

To explore the ability of the MMF model in item cold-start situations, we test the MMF model and MF model on the datasets that simulate item cold-start situation. We use RMSE to measure the performance as well.

\begin{figure}[h]
\centerline{\includegraphics[width=0.40\textwidth]{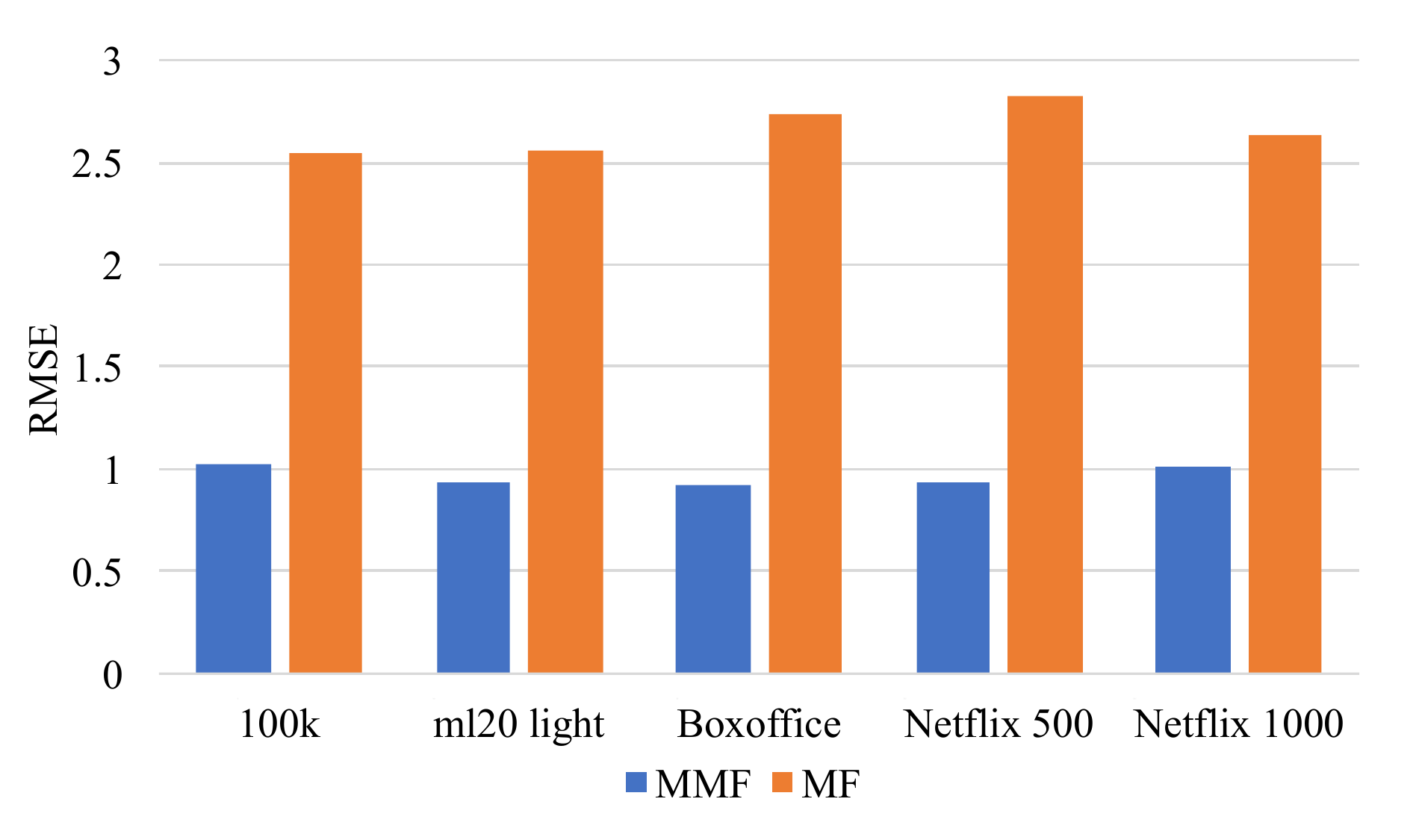}}
\caption{Prediction results on item cold-start situation.}
\label{Figcs}
\end{figure}

Figure \ref{Figcs} shows the RMSE results of two models on item cold-start situation. We can see that the MMF model has significant improvement on accuracy comparing to the MF model. It indicates that the MMF model successfully leverages attributes in items and gives accurate recommendation predictions on the movies without rating history. The MF model, however, provides nearly random predictions. Since the item cold-start problem happens frequently in practice, the MMF model shows greater value on practical applications.  

\subsection{Evaluation on Different Latent Vector Lengths}

In this subsection, we evaluate the performance of state-of-the-art matrix factorization models and MMF on different latent vector length. We test 4 different latent vector length (5, 10 ,15, 20) in this experiment.

\begin{figure*}[ht]
\centerline{\includegraphics[width=0.8\textwidth, height=5.4cm]{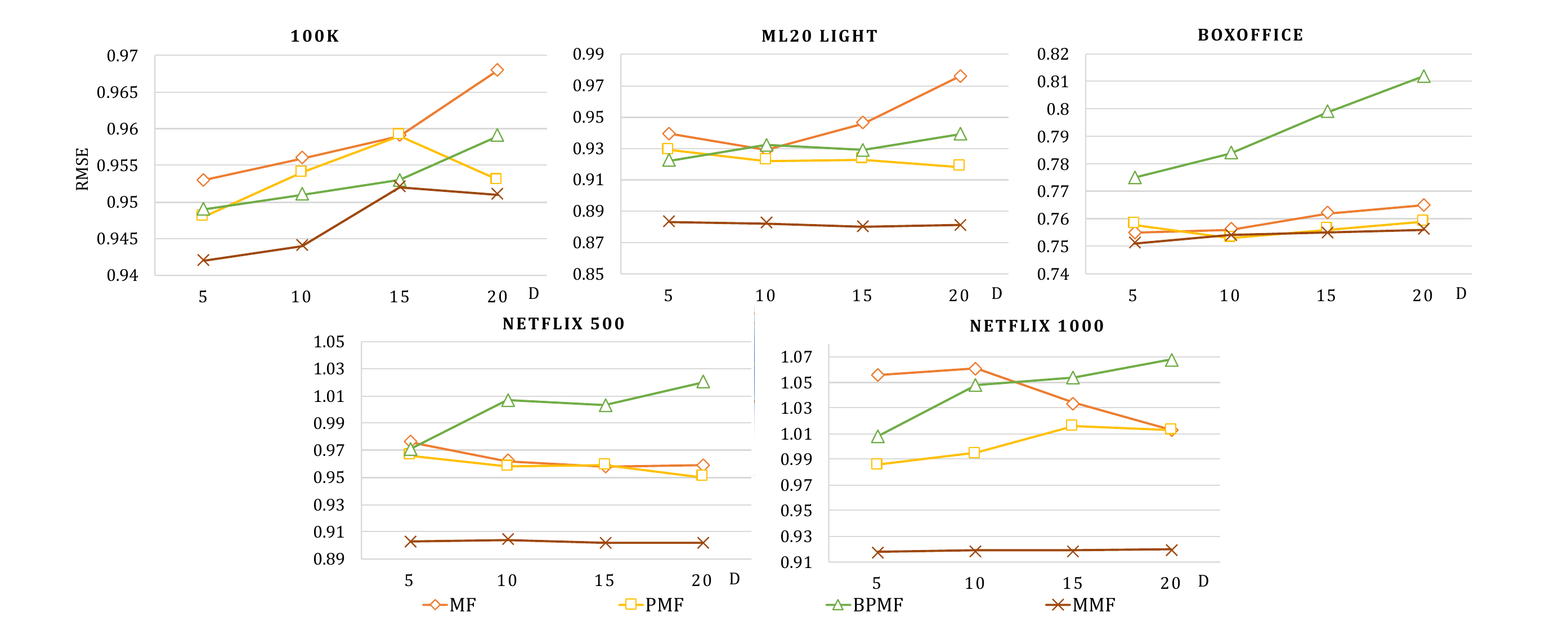}}
\caption{Test results of different latent vector length on MMF and baseline methods.}
\label{figlength}
\end{figure*}

Figure \ref{figlength} shows the test results. From the figure, we can see that for all datasets, MMF has better RMSE in all length of latent vectors. Moreover, the performance of MMF is more consistent than other methods when vector length changes. This indicates that our has more robust performance in different lengths of latent vectors.

\subsection{Evaluation on Different Topic Numbers} \label{sec_topicnum}

In previous experiments, we fix the topic number to 50 for all datasets. In this part, we evaluate our model's prediction accuracy on different topic numbers. We set 5 different topic numbers for all datasets, they are 0 (ignore topic attribute), 10, 20, 50 and 100. The results are shown in Table \ref{tab-topic}.

\begin{table}[htbp]
\small
\caption{RMSE of different topic numbers.}
\begin{center}
\begin{tabular}{|c|c|c|c|c|c|}
\hline
\textbf{Dataset}&\textbf{0}&\textbf{10}&\textbf{20}&\textbf{50}&\textbf{100} \\
\hline
100k & 0.942 & 0.938 & 0.938 & 0.936 & 0.935 \\
\hline
20m light & 0.889 & 0.877 & 0.880 & 0.876 & 0.875\\
\hline
Boxoffice & 0.757 & 0.751 & 0.753 & 0.750 & 0.751\\
\hline
Netflix 500 & 0.908 & 0.903 & 0.904 & 0.902 & 0.903\\
\hline
Netflix 1000 & 0.923 & 0.921 & 0.919 & 0.919 & 0.920\\
\hline
\end{tabular}
\label{tab-topic}
\end{center}
\end{table}

From the table, we find that considering topic attribute (topic number is larger than 0) improves the performance than ignoring it. It demonstrate that the topic information is beneficial to the MMF model's performance improvement. However, among the last four columns, there are not much difference on the performance when topic number changes. That may because that the topic attribute are generated from LDA model, which may not be accurate. Adding topic number may not provide with more useful information. Moreover, it also proves that MMF is robust: even the attributes are not accurate enough, the MMF model still performs firm and steady. 

\section{Conclusion and Future Work} \label{sec_conclusion}

In this work, we have proposed a nodel attribute interpretable collaborative filtering model called \textit{Multi-Matrix Factorization} (MMF). It offers great attribute level interpretability in the sense that it explains the contributions of each attribute affects the overall predicted rating, and at the same time it solves the item cold-start problem, which are two important limitations of matrix factorization model. Meanwhile, MMF outperforms state-of-the-art baselines in terms of accuracy. In the future, we will try MMF on datasets in other domains than movies to further validate the effectiveness. We will also explore the potential of embedding more powerful attribute analysis techniques to further improve the prediction accuracy. 

\section*{Acknowledgments}
This work is supported by China Scholarship Council (CSC) under the Grant CSC \#201808240005, and Australian Research Council (ARC) Discovery Project DP180102050. 

%Financial support from the program of China Scholarships Council (CSC) \#201808240005.

\bibliographystyle{IEEEtran}
\bibliography{IEEEabrv,reference}

\begin{thebibliography}{10}
\providecommand{\url}[1]{#1}
\csname url@samestyle\endcsname
\providecommand{\newblock}{\relax}
\providecommand{\bibinfo}[2]{#2}
\providecommand{\BIBentrySTDinterwordspacing}{\spaceskip=0pt\relax}
\providecommand{\BIBentryALTinterwordstretchfactor}{4}
\providecommand{\BIBentryALTinterwordspacing}{\spaceskip=\fontdimen2\font plus
\BIBentryALTinterwordstretchfactor\fontdimen3\font minus
  \fontdimen4\font\relax}
\providecommand{\BIBforeignlanguage}[2]{{%
\expandafter\ifx\csname l@#1\endcsname\relax
\typeout{** WARNING: IEEEtran.bst: No hyphenation pattern has been}%
\typeout{** loaded for the language `#1'. Using the pattern for}%
\typeout{** the default language instead.}%
\else
\language=\csname l@#1\endcsname
\fi
#2}}
\providecommand{\BIBdecl}{\relax}
\BIBdecl

\bibitem{wang2018kdgan}
X.~Wang, R.~Zhang, Y.~Sun, and J.~Qi, ``Kdgan: Knowledge distillation with
  generative adversarial networks,'' in \emph{Advances in Neural Information
  Processing Systems (NeurIPS)}, 2018, pp. 783--794.

\bibitem{davidson2010youtube}
J.~Davidson, B.~Liebald, J.~Liu, P.~Nandy, T.~Van~Vleet, U.~Gargi, S.~Gupta,
  Y.~He, M.~Lambert, B.~Livingston \emph{et~al.}, ``The youtube video
  recommendation system,'' in \emph{Proceedings of the fourth ACM Conference on
  Recommender Cystems (RecSys)}, 2010, pp. 293--296.

\bibitem{02sharma2013survey}
L.~Sharma and A.~Gera, ``A survey of recommendation system: Research
  challenges,'' \emph{International Journal of Engineering Trends and
  Technology (IJETT)}, vol.~4, no.~5, pp. 1989--1992, 2013.

\bibitem{liu2017cpmf}
C.-Y. Liu, C.~Zhou, J.~Wu, H.~Xie, Y.~Hu, and L.~Guo, ``Cpmf: A collective
  pairwise matrix factorization model for upcoming event recommendation,'' in
  \emph{2017 International Joint Conference on Neural Networks (IJCNN)}, 2017,
  pp. 1532--1539.

\bibitem{wang2018joint}
X.~Wang, J.~Qi, K.~Ramamohanarao, Y.~Sun, B.~Li, and R.~Zhang, ``A joint
  optimization approach for personalized recommendation diversification,'' in
  \emph{Pacific-Asia Conference on Knowledge Discovery and Data Mining
  (PAKDD)}.\hskip 1em plus 0.5em minus 0.4em\relax Springer, 2018, pp.
  597--609.

\bibitem{04shashua2005non}
A.~Shashua and T.~Hazan, ``Non-negative tensor factorization with applications
  to statistics and computer vision,'' in \emph{Proceedings of the 22nd
  International Conference on Machine Learning (ICML)}, 2005, pp. 792--799.

\bibitem{05xu2003document}
W.~Xu, X.~Liu, and Y.~Gong, ``Document clustering based on non-negative matrix
  factorization,'' in \emph{Proceedings of the 26th International ACM
  Conference on Research and Development in Informaion Retrieval (SIGIR)},
  2003, pp. 267--273.

\bibitem{06vlachos2016toward}
M.~Vlachos, V.~G. Vassiliadis, R.~Heckel, and A.~Labbi, ``Toward interpretable
  predictive models in b2b recommender systems,'' \emph{IBM Journal of Research
  and Development}, vol.~60, no. 5/6, pp. 11--1, 2016.

\bibitem{07lam2008addressing}
X.~N. Lam, T.~Vu, T.~D. Le, and A.~D. Duong, ``Addressing cold-start problem in
  recommendation systems,'' in \emph{Proceedings of the 2nd International
  Conference on Ubiquitous Information Management and Communication (IMCOM)},
  2008, pp. 208--211.

\bibitem{08zhang2016collaborative}
F.~Zhang, N.~J. Yuan, D.~Lian, X.~Xie, and W.-Y. Ma, ``Collaborative knowledge
  base embedding for recommender systems,'' in \emph{Proceedings of the 22nd
  International ACM Conference on Knowledge Discovery and Data Mining
  (SIGKDD)}, 2016, pp. 353--362.

\bibitem{09wang2011collaborative}
C.~Wang and D.~M. Blei, ``Collaborative topic modeling for recommending
  scientific articles,'' in \emph{Proceedings of the 17th ACM International
  Conference on Knowledge Discovery and Data Mining (SIGKDD)}, 2011, pp.
  448--456.

\bibitem{wang2015collaborative}
H.~Wang, N.~Wang, and D.-Y. Yeung, ``Collaborative deep learning for
  recommender systems,'' in \emph{Proceedings of the 21th ACM International
  Conference on Knowledge Discovery and Data Mining (SIGKDD)}, 2015, pp.
  1235--1244.

\bibitem{cheng2016wide}
H.-T. Cheng, L.~Koc, J.~Harmsen, T.~Shaked, T.~Chandra, H.~Aradhye,
  G.~Anderson, G.~Corrado, W.~Chai, M.~Ispir \emph{et~al.}, ``Wide \& deep
  learning for recommender systems,'' in \emph{Proceedings of the 1st Workshop
  on Deep Learning for Recommender Systems (RecSys)}, 2016, pp. 7--10.

\bibitem{shan2016deep}
Y.~Shan, T.~R. Hoens, J.~Jiao, H.~Wang, D.~Yu, and J.~Mao, ``Deep crossing:
  Web-scale modeling without manually crafted combinatorial features,'' in
  \emph{Proceedings of the 22nd ACM International Conference on Knowledge
  Discovery and Data Mining (SIGKDD)}, 2016, pp. 255--262.

\bibitem{11zhang2014explicit}
Y.~Zhang, G.~Lai, M.~Zhang, Y.~Zhang, Y.~Liu, and S.~Ma, ``Explicit factor
  models for explainable recommendation based on phrase-level sentiment
  analysis,'' in \emph{Proceedings of the 37th ACM International Conference on
  Research and Development in Information Retrieval (SIGIR)}, 2014, pp. 83--92.

\bibitem{wang2018explainable}
N.~Wang, H.~Wang, Y.~Jia, and Y.~Yin, ``Explainable recommendation via
  multi-task learning in opinionated text data,'' in \emph{Proceedings of the
  41th ACM International Conference on Research and Development in Information
  Retrieval (SIGIR)}, 2018, pp. 165--174.

\bibitem{18rendle:tist2012}
S.~Rendle, ``Factorization machines with {libFM},'' \emph{ACM Transactions on
  Intelligent Systems and Technology (TIST)}, vol.~3, no.~3, pp. 57:1--57:22,
  2012.

\bibitem{he2017neural}
X.~He, L.~Liao, H.~Zhang, L.~Nie, X.~Hu, and T.-S. Chua, ``Neural collaborative
  filtering,'' in \emph{Proceedings of the 26th International Conference on
  World Wide Web (WWW)}, 2017, pp. 173--182.

\bibitem{ortega2016recommending}
F.~Ortega, A.~Hernando, J.~Bobadilla, and J.~H. Kang, ``Recommending items to
  group of users using matrix factorization based collaborative filtering,''
  \emph{Information Sciences}, vol. 345, pp. 313--324, 2016.

\bibitem{ZhengTDZ16}
Y.~Zheng, B.~Tang, W.~Ding, and H.~Zhou, ``A neural autoregressive approach to
  collaborative filtering,'' in \emph{Proceedings of the 33nd International
  Conference on Machine Learning (ICML)}, 2016, pp. 764--773.

\bibitem{12harper2016movielens}
F.~M. Harper and J.~A. Konstan, ``The movielens datasets: History and
  context,'' \emph{ACM Transactions on Interactive Intelligent Systems (TIIS)},
  vol.~5, no.~4, p.~19, 2016.

\bibitem{13bennett2007netflix}
J.~Bennett, S.~Lanning \emph{et~al.}, ``The netflix prize,'' in
  \emph{Proceedings of Knowledge Discovery and Data Minming Cup and Workshop
  (SIGKDD)}, 2007, p.~35.

\bibitem{15blei2003latent}
D.~M. Blei, A.~Y. Ng, and M.~I. Jordan, ``Latent dirichlet allocation,''
  \emph{Journal of Machine Learning Research}, vol.~3, no. Jan, pp. 993--1022,
  2003.

\bibitem{16mnih2008probabilistic}
A.~Mnih and R.~R. Salakhutdinov, ``Probabilistic matrix factorization,'' in
  \emph{Processings of Advances in Neural Information Processing Systems
  (NIPS)}, 2008, pp. 1257--1264.

\bibitem{17salakhutdinov2008bayesian}
R.~Salakhutdinov and A.~Mnih, ``Bayesian probabilistic matrix factorization
  using markov chain monte carlo,'' in \emph{Proceedings of the 25th
  International Conference on Machine Learning (ICML)}, 2008, pp. 880--887.

\bibitem{maaten2008visualizing}
L.~v.~d. Maaten and G.~Hinton, ``Visualizing data using t-sne,'' \emph{Journal
  of machine learning research (JMLR)}, vol.~9, no. Nov, pp. 2579--2605, 2008.

\end{thebibliography}

\end{document}